# Elastic and Inelastic Electron Scattering Cross Sections of Trichlorofluoromethane

M. Dinger[1,2,*], Y. Park[3], and W. Y. Baek[1]

[1]Physikalisch-Technische Bundesanstalt (PTB), Bundesallee 100, 38116 Braunschweig, Germany

[2]Ruprecht-Karls-Universität Heidelberg, Grabengasse 1, 69117 Heidelberg, Germany

[3]Institute of Plasma Technology, Korea Institute of Fusion Energy (KFE), 37, Dongjangsan-ro, Gunsan, Jeonbuk-do 54004, Republic of Korea

[*]Contact author: mareike.dinger@ptb.de

**Abstract** Differential elastic electron scattering cross sections of trichlorofluoromethane ($CCl_3F$) were measured for the first time over a broad energy range spanning 30 eV to 800 eV in the angular range of 20° to 150°. The experimental results were compared with calculations using the IAM-SCAR+I model. Satisfactory agreements between both data sets were found for electron energies above 200 eV within experimental uncertainties, whereas significant deviations of up to 100% were observed at electron energies below 60 eV. In addition to the measurements of differential elastic scattering cross sections, total inelastic scattering cross sections of $CCl_3F$ were calculated using the spherical complex optical potential (SCOP) model. These calculations closely match experimental total ionization cross sections available in the literature for energies below 50 eV. The sum of the experimental total elastic and the theoretical total inelastic scattering cross sections align very well with the total electron scattering cross sections of $CCl_3F$ measured by other groups across the entire energy range (30 eV to 800 eV), demonstrating the consistency among these three cross sections.

## 1. Introduction

Chlorofluorocarbons (CFCs) were the primary choice for refrigeration and various industrial processes until they were banned by the Montreal Protocol due to their ozone-depleting effects. Among these, trichlorofluoromethane ($CCl_3F$, Freon-11, CFC-11) has a particular notorious impact on chemical processes in the upper atmosphere due to the presence of three chlorine atoms, leading to a global warming potential several orders of magnitude higher than that of $CO_2$ [1].

There is clear evidence that cosmic ray-driven electron-induced molecular reactions play an important role in ozone depletion [2]. Electron impact on $CCl_3F$ results in the production of Cl radicals, which catalyze ozone depletion. With an atmospheric lifetime of 52 years and a high ozone depletion potential [3], understanding the atmospheric chemistry of $CCl_3F$ is crucial. Accurate simulations of radiation transport processes in the upper atmosphere rely heavily on the availability of data sets on electron-molecule collisions.

Hitherto, only a few studies on the electron-impact cross sections of $CCl_3F$ have been conducted. The first measurement of the electron interaction cross sections of $CCl_3F$ dates back to 1986, when Jones [4] determined the total electron scattering cross sections (TCS) with a time-of-flight electron transmission spectrometer in the energy range from 0.6 eV to 50 eV. A few years later, Zecca et al. [5] extended this range by providing experimental TCS for electron energies between 75 eV and 4 keV. Jiang et al. [6] later calculated ionization cross sections using the binary-encounter Bethe model (BEB) for energies from 10 eV to 1 keV. There is only one experimental study on the total ionization cross sections of $CCl_3F$. Sierra et al. [7] measured the partial ionization cross sections of $CCl_3F$ by collecting the fragment ions upon electron-impact for primary energies between 20 eV and 85 eV. The total ionization cross section was then obtained by summing the partial ionization cross sections. Martinez et al. [8] contributed to this study by comparing the experimental results with theoretical calculations based on the BEB, Deutsch and Märk formalism, and the modified additivity rule.

Despite the importance of electron collision cross sections of $CCl_3F$ for simulating radiation transport processes, only one experimental dataset on the differential elastic electron scattering cross sections (DCS) of $CCl_3F$ is currently available. In this dataset, Hoshino et al. [9] not only provided experimental DCS in the angular range of 15°-130°, but also discussed the role of the chlorine atoms by comparing their results to those of other CFCs. Additionally, they compared their experimental DCS to calculations performed using the IAM-SCAR model and a modified phase shift analysis (PSA). For electron energies up to 30 eV, further theoretical DCS data were reported by Natalese et al. [10] and Freitas et al. [11], who employed the Schwinger multichannel method for their calculations. However, all available literature on DCS focuses solely on energies below 100 eV. To provide a comprehensive dataset for radiation transport calculations, we measured the DCS $\mathrm{d}\sigma_{\mathrm{el}}/\mathrm{d}\Omega$ of $CCl_3F$ for the first time over a broader energy and angular range, covering scattering angles $\theta$ from 20° to 150° and electron energies $T$ between 30 eV and 800 eV. The experimental results were compared to calculations using the IAM-SCAR+I model [12–14]. Based on the experimental DCS, the total elastic scattering cross sections (TECS) $\sigma_{\mathrm{el}}$ and the momentum transfer cross sections (MTCS) $\sigma_{\mathrm{m}}$ were determined and compared to the calculations with the IAM-SCAR+I model and the close-coupling code POLYDCS [15]. Additionally, the total inelastic scattering cross section (TICS) $\sigma_{\mathrm{inel}}$ of $CCl_3F$ was calculated using the spherical complex optical potential (SCOP) model [16] and compared to the experimental total ionization cross sections reported by Sierra et al. [7]. To check the consistency and reliability of both the experimental and theoretical cross sections, the sum of TECS and TICS was compared to the experimental data for the TCS $\sigma_{\mathrm{tot}}$ of $CCl_3F$.

## 2. Experiment

The DCS of $CCl_3F$ was measured using a crossed-beam setup, as depicted in Fig. 1. Since the experimental setup was thoroughly explained in our previous works [17], we provide only a brief summary here. In this crossed-beam setup, the primary electron beam intersects perpendicularly with a molecular beam created by an effusion nozzle. The energy spectrum of scattered electrons is measured with a hemispherical energy analyzer mounted on a turntable. The scattering angle $\theta$ is adjusted by rotating the turntable, while the electron gun remains fixed. Since the detection solid angle is nearly constant throughout the interaction zone in the present experimental setup, the DCS can be obtained from the change in the elastic count rate $\Delta\dot{N}_{\mathrm{el}}$ per solid angle $\Delta\Omega$:

$$\frac{d\sigma_{\mathrm{el}}}{d\Omega}(\theta,T) = \frac{\Delta\dot{N}_{\mathrm{el}}}{\Delta\Omega}(\theta,T)\Big/\left(\frac{-I_0}{e}n_{\mathrm{F}}\eta(T)\right), \tag{1}$$

where $I_0$ is the primary electron current (negative in sign), $e$ is the elementary charge, $n_{\mathrm{F}}$ is the number of molecules per area hit by the electron beam, and $\eta(T)$ is the detection efficiency of the energy analyzer.

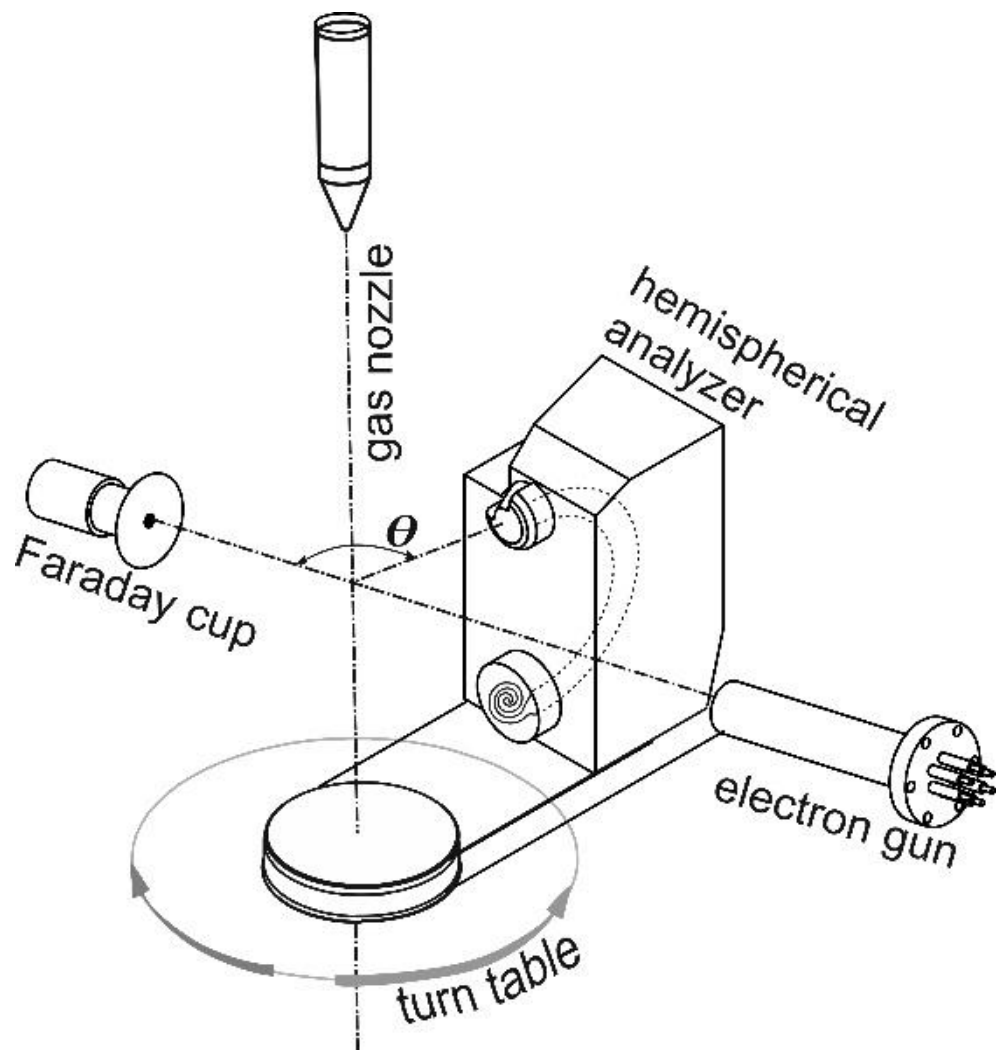

FIG. 1. Schematic view of the experimental setup. The electron gun and Faraday cup were fixed in position, while the hemispherical electron energy analyzer was mounted on a turntable. The scattering angle $\theta$ of electrons to be detected was adjusted by rotating the turntable.

Direct determination of $n_{\mathrm{F}}$ and $\eta(T)$ is challenging. In this work, this difficulty is circumvented by using the relative flow technique (RFT) [18], which utilizes the well-known DCS of a reference gas. To use the RFT, the molecular beams must be generated in the molecular flow regime, and the mean free paths for intermolecular collisions in both the reference gas and the gas of interest must be comparable. If these conditions are met, the ratio $n_{\mathrm{F}}/\hat{n}_{\mathrm{F}}$ of the area number densities in the two beams is given by

$$\Gamma \equiv n_{\mathrm{F}}/\hat{n}_{\mathrm{F}} = \hat{F}/F \times \sqrt{M/\hat{M}}, \tag{2}$$

where $F$ is the mass flow rate through the effusion nozzle and $M$ is the mass of the molecule of interest. In Eq. (2) and the following equations, the quantities with hats denote those related to the reference gas. With $\Gamma$, Eq. (1) can be reformulated as

$$\frac{d\sigma_{\mathrm{el}}}{d\Omega}(\theta,T) = \frac{d\hat{\sigma}_{\mathrm{el}}}{d\Omega}(\theta,T) \times \left(\frac{\Delta\dot{N}_{\mathrm{el}}}{\Delta\Omega}\Big/\frac{\Delta\hat{\dot{N}}_{\mathrm{el}}}{\Delta\Omega}\right) \times \frac{\hat{I}_0}{I_0} \times \Gamma^{-1}. \tag{3}$$

In the present work, nitrogen ($N_2$) was used as the reference gas, with its DCS values determined in an independent prior experiment that utilized a different methodology and helium as the reference gas [17]. Although numerous experimental DCS datasets for $N_2$ have been published to date [19–26], these data are restricted to specific angular and energy ranges, and none cover the broad energy and angular ranges considered in this study. Moreover, the published data exhibit significant scatter, with deviations exceeding 50% in some cases. To establish a suitable reference dataset for this study, the existing data would need to be interpolated and extrapolated to encompass the scattering angles and primary electron energies relevant to this work, introducing additional uncertainties.

At the scattering angles and energies studied in the present work, our reference data generally align with the mean literature values within our experimental uncertainties of approximately 15% and the standard deviations of the average literature data. When only two or fewer datasets were available, we also considered the stated experimental uncertainties from the literature. A notable deviation of 114% was found at 135° for $T = 60$ eV between our reference DCS and the literature data, which cannot be solely attributed to experimental uncertainties. This discrepancy originates from the dataset of Srivastava et al. [22], which is not only higher than our own reference data but also exceeds the only other available dataset at this energy (Nickel et al. [25]). Apart from this outlier, our reference $N_2$ data are consistent with both the averaged literature values [19–25] and the compiled reference dataset of Song et al. [26] within experimental uncertainties. Therefore, the use of our $N_2$ DCS measurements as the reference dataset is justified and appropriate for this study.

The number density of molecules in the gas beam, produced via an effusion nozzle with a 0.3 mm diameter exit aperture, was approximately $5 \times 10^{13}$ cm$^{-3}$. Considering the gas kinetic diameters of $N_2$ (0.364 nm [27]) and $CCl_3F$ (0.618 nm [28]), the mean free paths ($\lambda$) for intermolecular collisions in both gases exceed 4 mm. This is significantly larger than the nozzle diameter, ensuring the Knudsen condition is well satisfied. Consequently, the gas effusion occurs in the molecular flow regime, allowing for the application of the RFT. To maintain equal $\lambda$ in both gas beams, a necessary condition for RFT, the driving pressure in the gas reservoir above the effusion nozzle was adjusted according to the square of the gas kinetic diameters of both molecules. Given that the total electron scattering cross sections of $N_2$ and $CCl_3F$ for the energies of interest are lower than $6\times10^{-15}$ cm$^2$, the mean free path of electrons in both gas beams with the aforementioned number density exceeded 3 cm, which was much larger than the diameter of the molecular beam. Consequently, the single collision condition was satisfied.

The primary electron beam current, measured using the Faraday cup located beyond the molecular beam, varied between 0.2 μA and 0.5 μA. The angular resolution of the apparatus was approximately 2°. The overall energy resolution of the apparatus was 1.5 eV at $T$= 800 eV, improving to 0.7 eV at $T$= 30 eV. It should be noted that this energy resolution is not sufficient to resolve rotational and certain vibrational excitations from elastic scattering. The scattering chamber was surrounded by three orthogonal pairs of Helmholtz coils to compensate for the Earth's magnetic field. On the scattering plane, the residual magnetic field was lower than 2 μT. The deflection of electrons in this field was negligible compared to the angular resolution of the apparatus and was therefore disregarded.

## 3. Uncertainty analysis

The standard uncertainty of the experimental DCS was estimated following the Guide to the Expression of Uncertainty in Measurement [29]. The major uncertainty source is the DCS of the reference gas $N_2$, which had 15% uncertainty. Another uncertainty arose from the temporal drift of the primary electron beam current, which was corrected with an uncertainty of 4%. The ratio of the flow rates of the $CCl_3F$ and $N_2$ gas beams, determined from the temporal decrease of the pressure in their gas reservoirs, was associated with a 5% uncertainty. Finally, the statistical uncertainties due to the scattered electron counts under the elastic peaks were lower than 5%. The overall standard uncertainty, calculated as the positive square root of the sum of the squared individual uncertainties, was 17%.

## 4. Theoretical Methods

Unless otherwise stated, the static $V_{\mathrm{st}}$, exchange $V_{\mathrm{ex}}$ and correlation-polarization potential $V_{\mathrm{cp}}$ as well as the electron densities $\varrho$ needed in the following theories were obtained in their single-center expansion using the SCELIB4.0 library [30]. The molecular wavefunctions used as input in the SCELib4.0 library were computed using the Hartree-Fock method with the GAUSSIAN09 [31] program suite. A 6-311++G basis set, supplemented with additional (2d,p) polarization functions was employed for these calculations. The optimized structure and polarizabilities are summarized in Tab. 1. Within SCELIB4.0, the correlation-polarization potential $V_{\mathrm{cp}}$ was calculated using the modified free electron gas model, and the exchange potential was obtained using the Hara Free Electron Gas Exchange (HFEGE) model, with detailed information provided in Ref. [30].

Figure 2 illustrates these potentials for $T$ = 100 eV. While $V_{\mathrm{st}}$ and $V_{\mathrm{cp}}$ are independent of electron energy, $V_{\mathrm{ex}}$ increases as the electron energy decreases. The peak at $r$ = 0.475 a.u. arises from the carbon atom, whereas the one around $r$ = 3.0 a.u. is associated with the chlorine and fluorine atoms.

**TABLE 1** Molecular geometry and polarizability of $CCl_3F$ determined through Hartree-Fock optimization using the GAUSSIAN09 program. The atomic coordinates $x$, $y$ and $z$ are given in units of Å and the polarizability tensor elements $\alpha_{ij}$ are provided in atomic units. The center of mass was chosen as the origin.

| | x | y | z | $\alpha$ | |
|---|---|---|---|---|---|
| C | 0.000000 | 0.000000 | 0.251350 | $\alpha_{xx}$ | 54.043 |
| Cl | 0.000000 | 1.672954 | -0.305277 | $\alpha_{yy}$ | 54.043 |
| Cl | 1.448821 | 0.836477 | -0.305277 | $\alpha_{zz}$ | 39.470 |
| Cl | 1.448821 | 0.836477 | -0.305277 | $\alpha_{xz}, \alpha_{yz}, \alpha_{xy}$ | 0.000 |
| F | 0.000000 | 0.000000 | 1.562337 | | |

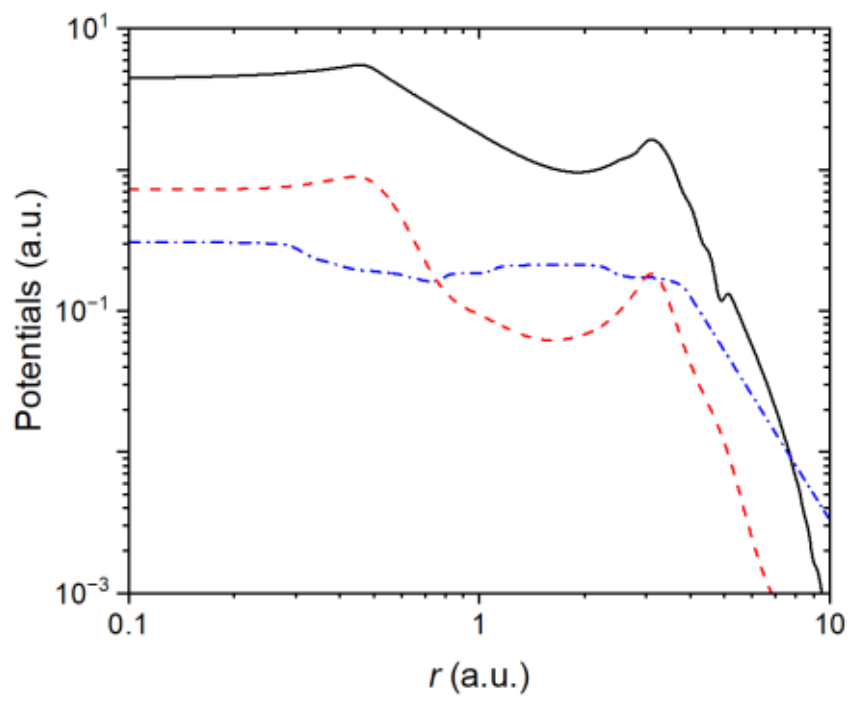

FIG. 2. Potentials used for the calculation in the IAM-SCAR+I und SCOP model in atomic units for $T = 100$ eV: $V_{\mathrm{st}}$ (—), $V_{\mathrm{ex}}$ (- - -), $V_{\mathrm{cp}}$ (– · –).

### 4.1. Elastic scattering

In first order, the elastic DCS of molecules can be determined using the independent atom model with additivity rule (IAM-AR). In this approach, the molecular DCS is obtained by coherently summing the squared scattering amplitudes $f_i(\theta)$ of each atomic constituent (C, Cl, F). The elastic scattering amplitudes are derived from the interaction potentials of the atomic constituents, which include i) a static term, ii) an exchange term, and iii) a term characterizing long-range interactions arising from the molecular polarizability. While the IAM-AR provides a useful approximation at high energies, it neglects screening and interference effects arising from different scattering centers in polyatomic molecules, which become increasingly important at lower electron energies.

Blanco and García further improved this approach in their IAM-SCAR+I model [12–14,32] by incorporating screening and interference effects based on the molecular geometry:

$$\left(\frac{\mathrm{d}\sigma_{\mathrm{el}}}{\mathrm{d}\Omega}\right)^{\mathrm{direct}} = \sum_{i}^{N} s_i^2 |f_i(\theta)|^2 + \sum_{i \neq j}^{N} v_{ij} s_i s_j f_i(\theta) f_j^*(\theta) \frac{\sin(q x_{ij})}{q x_{ij}}\,, \tag{4}$$

where $s_i$ and $v_{ij}$ are the screening and interference coefficients derived from the spatial arrangement of the constituent atoms. The terms $q$ an $x_{ij}$ describe the momentum transfer and distance between atoms $i$ and $j$, respectively. In addition to direct scattering described by Eq. (4), the IAM-SCAR+I model accounts for redispersed contributions arising from multiple-scattering processes. For detailed derivations and discussions of all terms, we refer readers to the original works [12–14,32]. In calculating the DCS with the IAM-SCAR+I model, the scattering amplitudes of the atomic and polarization potentials were determined following methods described in an earlier work [17], utilizing the Fermi model for the exchange potential and the modified free electron gas model [30] for the polarization potential.

Rotational and total elastic scattering cross sections (TECS) were determined using the close-coupling code POLYDCS [15] with the required K-matrix obtained through the VOLSCAT package [33]. In the VOLSCAT code, the K-matrix is calculated by numerically integrating the integro-differential scattering equation, with the procedure governed by input parameters defining the interaction potentials, the projectile properties, and integration grids. As previously mentioned, the interaction potentials were derived using the SCELIB4.0 library. The numerical integration was performed on a grid with 32 polar and azimuthal points and 256 points along the radial coordinate, extending up to 250 atomic units. The K-matrix was computed up to a maximal angular momentum quantum number of 25, achieving a balance between computational efficiency and desired accuracy of the results. Using this K-matrix, the TECS, including rotational excitation cross sections of $CCl_3F$, were obtained via the POLYDCS code.

The main input parameters for POLYDCS include rotational constants, the molecular dipole moment, rotational state quantum numbers, and the maximum angular momentum quantum number. The rotational constants along the principal axes are 0.08203 $cm^{-1}$, 0.08203 $cm^{-1}$, and 0.05738 $cm^{-1}$ [34], while the dipole moment of $CCl_3F$ is 0.46 D [35]. The total rotational excitation cross sections were determined by summing the individual rotational excitation cross sections for transitions from the ground state up to the 9th excited state, with the molecule modeled as an asymmetric rigid rotor. Differential elastic scattering cross sections in the POLYDCS code are expressed as Legendre expansions, with the maximum number of terms set to 50 for this study.

### 4.2. Inelastic scattering

In the spherically complex optical potential (SCOP) model, electron scattering is represented by a complex interaction potential $V(\vec{r})$ consisting of both a real and an imaginary part:

$$V(\vec{r}) = V_{\mathrm{R}}(\vec{r}) + iV_{\mathrm{abs}}(\vec{r}), \tag{5}$$

where the real part $V_{\mathrm{R}}(\vec{r})$ accounts for elastic scattering, and the imaginary component $V_{\mathrm{abs}}(\vec{r})$ describes the absorption of incident electron flux into interaction channels, leading to inelastic scattering events.

The main feature of the SCOP model is the use of spherically symmetric potentials to describe the electron-molecule interaction. In this work, the spherically symmetric optical potential $V^{\mathrm{opt}}(r)$ was obtained by fully averaging $V(\vec{r})$ given by Eq. (5) over all possible molecular orientations. To facilitate this averaging, $V(\vec{r})$ was expanded around the center of mass of the molecule using symmetry-adapted functions with the SCELIB4.0 library

$$V(\vec{r}) = \sum_{lm} V_{lm}(r) X_{lm}^{\mathrm{A}_1}(\theta, \varphi). \tag{6}$$

In Eq. (6), $r$ is the distance from the center of mass of the molecule and $X_{lm}^{\mathrm{A}_1}$ is the symmetry-adapted function for the totally symmetric irreducible representation (IR) $\mathrm{A}_1$ for the angular momentum $l$ and its component $m$. The latter can be represented as a linear combination of real spherical harmonics $S_{lm}(\theta, \varphi)$:

$$X_{lm}^{\mathrm{A}_1}(\theta, \varphi) = \sum_{l=-m}^{m} b_{lm}^{\mathrm{A}_1} S_{lm}(\theta, \varphi), \tag{7}$$

where the coefficients $b_{lm}^{\mathrm{A}_1}$ can be obtained from the character table of the IR $\mathrm{A}_1$. The averaging of $V(\vec{r})$ was performed by integrating the right-hand side of Eq. (6) over the three Euler angles and then dividing it by $8\pi^2$. Due to the orthogonality of spherical harmonics over the surface of a sphere, the integral of a single real spherical harmonic over Euler angles vanishes for $l \neq 0$, so that $V^{\mathrm{opt}}(r)$ is given by $V^{\mathrm{opt}}(r) = V_{00}/\sqrt{4\pi}$ (with $b_{00}^{\mathrm{A}_1} = 1$).

The real part $V_{\mathrm{R}}(\vec{r})$ of the interaction potential comprises the static $V_{\mathrm{st}}$, exchange $V_{\mathrm{ex}}$ and correlation-polarization $V_{\mathrm{cp}}$ potentials mentioned in the beginning and the absorption potential $V_{\mathrm{abs}}(\vec{r})$ was generated using the quasifree-scattering model described in detail by Staszeweska et al. [36,37]:

$$V_{\mathrm{abs}}(\vec{r}) = -\frac{1}{2}\varrho(\vec{r})\sqrt{2(T - V_{\mathrm{SE}})}\,\frac{4\pi}{5k_{\mathrm{F}}^3 T} H(\gamma)(Z_1 + Z_2 + Z_3). \tag{8}$$

Here, $\varrho(\vec{r})$ is the electron density per unit volume, $V_{\mathrm{SE}}$ is the sum of the static and exchange potentials, $k_{\mathrm{F}}$ is the Fermi momentum given by $k_{\mathrm{F}} = (3\pi^2\varrho)^{1/3}$, and $H(\gamma)$ is the Heaviside step function with $\gamma = k^2 + k_{\mathrm{F}}^2 - \alpha - \beta$, where $k$ is the momentum of incident electrons. Among different models for $\alpha$ and $\beta$, we chose $\alpha = k_{\mathrm{F}}^2 + 2I - V_{\mathrm{SE}}$ and $\beta = k_{\mathrm{F}}^2 - V_{\mathrm{SE}}$, where $I = 11.73\ \mathrm{eV}$ [38] is the ionization potential of the molecule. The terms $Z_1$, $Z_2$, and $Z_3$ are given by

$$Z_1 = \frac{5k_{\mathrm{F}}^3}{\alpha - k_{\mathrm{F}}^2},\quad Z_2 = -\frac{k_{\mathrm{F}}^3[5(k^2 - \beta) + 2k_{\mathrm{F}}^2]}{(k^2 - \beta)^2},\quad Z_3 = H(\tilde{\gamma})\frac{2\tilde{\gamma}^{5/2}}{(k^2 - \beta)^2} \tag{9}$$

with $\tilde{\gamma} = \alpha + \beta - k^2$.

Once the spherical complex optical potential was obtained, the radial Schrödinger equation was solved to determine the phase shift of scattered electrons by applying the variable phase approach [39,40]. In this approach, the real part $\varepsilon_l$ and the imaginary part $\eta_l$ of the phase shifts are obtained from two coupled first-order differential equations:

$$\frac{d\varepsilon_l}{dr} = -\frac{1}{k}\left[V_{\mathrm{R}}^{\mathrm{opt}}(X^2 - Y^2) - 2V_{\mathrm{abs}}^{\mathrm{opt}}XY\right],\quad \frac{d\chi_l}{dr} = -\frac{1}{k}\left[V_{\mathrm{abs}}^{\mathrm{opt}}(X^2 - Y^2) + 2V_{\mathrm{R}}^{\mathrm{opt}}XY\right] \tag{10}$$

with

$$X = \cosh\chi_l\,[\eta_l \sin\varepsilon_l - j_l \cos\varepsilon_l],\ Y = \sinh\chi_l\,[\eta_l \cos\varepsilon_l + j_l \sin\varepsilon_l], \tag{11}$$

where $j_l(kr)$ and $\eta_l(kr)$ are the usual Riccati-Bessel functions. Equation (10) was solved by integrating it using the fourth-order Runge-Kutta method [41]. The phase shifts $\varepsilon_l(kr)$ and $\chi_l(kr)$ for $r \to \infty$ are related to the *S* matrix via $S_l = \exp[2i(\varepsilon_l + i\chi_l)]$, which is further linked to the TICS $\sigma_{\mathrm{inel}}$ through the equation

$$\sigma_{\mathrm{inel}}(k) = \frac{\pi}{k^2}\sum_{l}(2l+1)[1-|S_l(k)|^2]. \quad (12)$$

## 5. Results and Discussion

The results of the present experiment are listed in Table 2 and displayed in Fig. 3, where they are compared to both the experimental data of Hoshino et al. [9] and theoretical calculations. Since trichlorofluoromethane is a polar molecule with a permanent dipole moment [27], it is prone to substantial rotational excitations. Due to the finite energy resolution of our experimental setup, it was not possible to fully separate pure elastic DCS from contributions due to rotational and vibrational excitations. To provide a more accurate comparison, the rotational excitation cross sections were included in the theoretical models. The theoretical values shown in Fig. 3 include the DCS calculated using the IAM-SCAR+I model with and without the inclusion of rotational excitation cross sections of the $CCl_3F$ molecule.

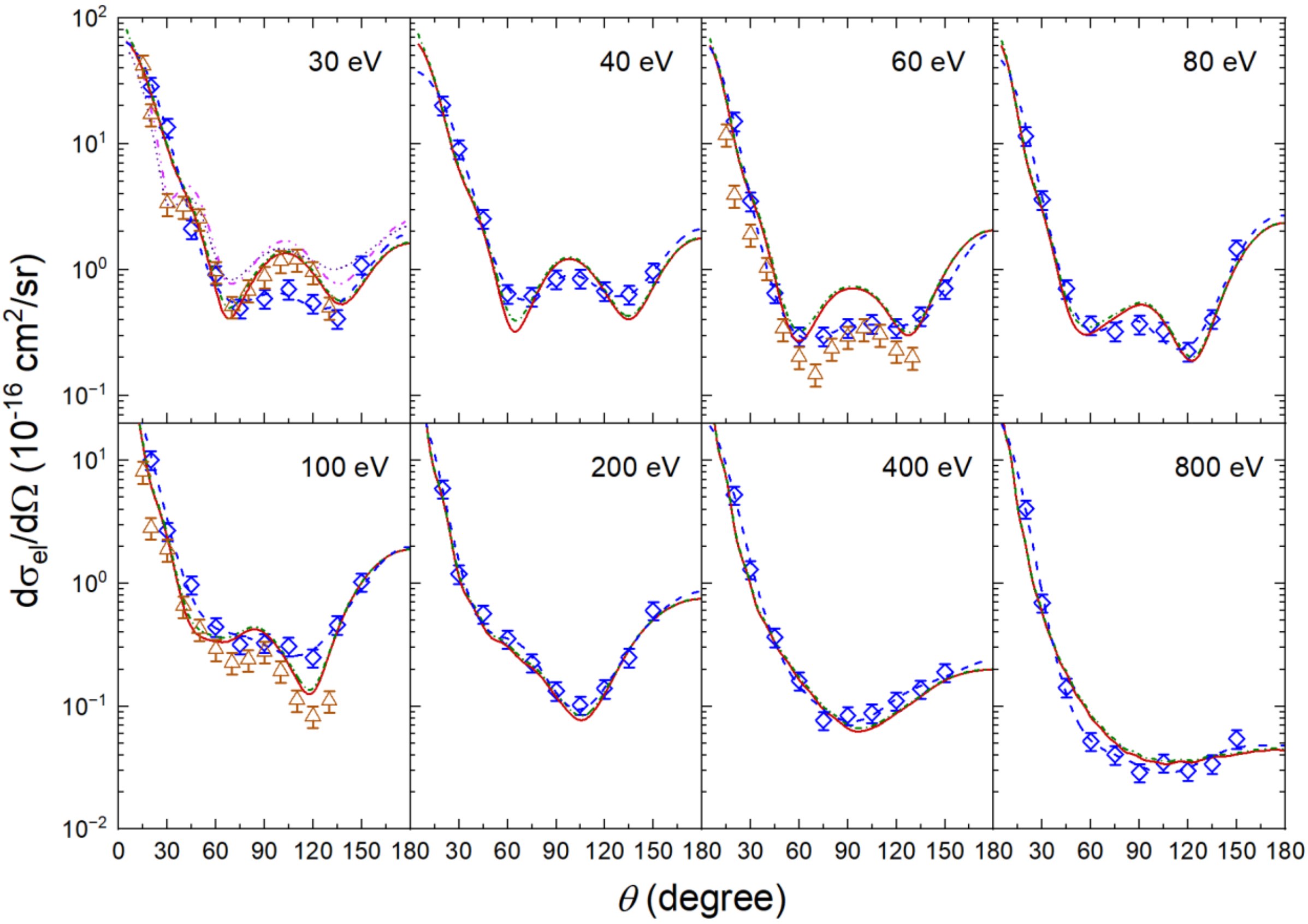

FIG. 3. Present experimental results (◇) for the DCS of $CCl_3F$ for different primary electron energies compared to the elastic DCS reported by Hoshino et al. [9] (△). The solid line (—) represents calculations using the IAM-SCAR+I model, the dash-dotted line (– · –) depicts calculations using the IAM-SCAR+I model plus the rotational excitation cross sections, and the dashed line (– –) describes the results of best fits of Eq. (13) to the experimental data. For 30 eV, Schwinger multichannel calculations with Born closure from Freitas et al. [11] were also available: static exchange (– · · –) and static exchange plus polarization approximations (· · ·).

It is important to note that the DCS data reported by Hoshino et al. [9] represent elastic scattering with only minor vibrational contributions, due to the experimental energy resolution of 35-40 meV, and are therefore expected to be slightly lower than both our experimental results and the theoretical values including rotational excitation cross sections.

TABLE 2. Present experimental results for the DCS of $CCl_3F$ as a function of the scattering angle $\theta$ for different electron energies $T$, expressed in units of $10^{-16}$ cm$^2$/sr. The overall uncertainties of the DCS are 17%. Additionally, the TECS $\sigma_{\rm el}$, MTCS $\sigma_{\rm m}$, TICS $\sigma_{\rm inel}$, and TCS $\sigma_{\rm tot}$ are given in units of $10^{-16}$ cm$^2$. The TICS was calculated using the SCOP model, and $\sigma_{\rm tot}$ was obtained by summing $\sigma_{\rm el}$ and $\sigma_{\rm inel}$.

| $\theta$ / $T$ | 30 eV | 40 eV | 60 eV | 80 eV |
|---|---|---|---|---|
| 20° | 28.4 | 20.2 | 15.1 | 11.5 |
| 30° | 13.5 | 9.09 | 3.50 | 3.61 |
| 45° | 2.10 | 2.53 | 0.653 | 0.705 |
| 60° | 0.908 | 0.642 | 0.297 | 0.366 |
| 75° | 0.494 | 0.615 | 0.298 | 0.324 |
| 90° | 0.589 | 0.843 | 0.348 | 0.370 |
| 105° | 0.700 | 0.849 | 0.372 | 0.326 |
| 120° | 0.539 | 0.676 | 0.345 | 0.224 |
| 135° | 0.408 | 0.636 | 0.431 | 0.405 |
| 150° | 1.09 | 0.963 | 0.709 | 1.46 |
| $\sigma_{\rm el}$ | $40.54 \pm 7.09$ | $31.75 \pm 5.37$ | $22.08 \pm 3.48$ | $20.09 \pm 3.46$ |
| $\sigma_{\rm m}$ | $11.58 \pm 2.03$ | $12.25 \pm 2.07$ | $7.41 \pm 1.17$ | $8.24 \pm 1.42$ |
| $\sigma_{\rm inel}$ | 7.21 | 9.16 | 10.43 | 10.35 |
| $\sigma_{\rm tot}$ | 47.75 | 40.91 | 32.51 | 30.44 |
| $\theta$ / $T$ | 100 eV | 200 eV | 400 eV | 800 eV |
| 20° | 10.1 | 5.91 | 5.25 | 4.05 |
| 30° | 2.67 | 1.20 | 1.30 | 0.698 |
| 45° | 0.973 | 0.566 | 0.365 | 0.143 |
| 60° | 0.444 | 0.354 | 0.162 | 0.0520 |
| 75° | 0.318 | 0.226 | 0.0770 | 0.0403 |
| 90° | 0.328 | 0.133 | 0.0840 | 0.0291 |
| 105° | 0.308 | 0.103 | 0.0882 | 0.0348 |
| 120° | 0.249 | 0.139 | 0.110 | 0.0299 |
| 135° | 0.461 | 0.250 | 0.139 | 0.0340 |
| 150° | 1.03 | 0.601 | 0.190 | 0.0545 |
| $\sigma_{\rm el}$ | $18.15 \pm 3.63$ | $10.51 \pm 2.24$ | $7.56 \pm 1.23$ | $5.49 \pm 1.49$ |
| $\sigma_{\rm m}$ | $7.16 \pm 1.43$ | $4.07 \pm 0.87$ | $2.05 \pm 0.33$ | $0.77 \pm 0.21$ |
| $\sigma_{\rm inel}$ | 10.89 | 9.57 | 7.41 | 5.14 |
| $\sigma_{\rm tot}$ | 29.04 | 20.08 | 14.97 | 10.63 |

The results of the present work, as shown in Fig. 3, agree satisfactorily with theoretical predictions for electron energies above 200 eV. However, for energies below 100 eV, notable discrepancies were observed when comparing our data to both the theory and the experimental data of Hoshino et al. [9]. These discrepancies are particularly evident in the angular range of 60° to 130°, where the DCS exhibit two distinct minima around 60° and 120°. While these minima are clearly observed in the experimental data of Hoshino et al. [9] as well as in the IAM-SCAR+I model, they are only faintly discernible in the present experimental data.

At 30 eV, the present results are generally lower than the DCS values reported by Hoshino et al. [9], which, except at 30°, are reasonably well reproduced by the IAM-SCAR+I model. The theoretical data reported by Freitas et al. [11] also show a similar angular dependence but significantly overestimate the DCS at the positions of the two minima. In terms of absolute values, the data from this work at 60 eV and 100 eV appear to align more closely with the theoretical values compared to the corresponding data from Hoshino et al. [9]. As noted above, the results of Hoshino et al., however, show better agreement with the theoretical predictions regarding the angular dependence of the DCS. It is important to mention that the calculated values have limited reliability at low energies due to the approximations used in the model, which are only roughly valid for electron energies below 50 eV. This limitation may partly explain the discrepancies observed at lower energies.

The double-minima structure observed in the experimental data, which varies with electron energy, is primarily attributed to the presence of the three chlorine atoms in $CCl_3F$. For example, the minimum around 120° is closely linked to the resonance-like sharp minima in the DCS of individual chlorine atoms (e.g., at $\theta = 118°$ and $T = 109.5$ eV [42]). These chlorine atoms dominate the DCS of $CCl_3F$ with an estimated contribution of approximately 75%. This phenomenon was also reported by Hoshino et al. [9], who found that the minimum around 120° became increasingly pronounced in the DCS as the number of chlorine atoms increased in $CF_{4-x}Cl_x$ molecules.

To obtain the TECS and the MTCS, the experimental DCS shown in Fig. 3 was fitted using a series of Legendre functions. In general, the DCS of a polyatomic molecule can be expressed by the following equation [15]:

$$\frac{\mathrm{d}\sigma_{\mathrm{el}}}{\mathrm{d}\Omega} = \frac{\mathrm{d}\sigma^{\mathrm{B}}}{\mathrm{d}\Omega} + \sum_{L}(A_L - A_L^{\mathrm{B}})P_L(\cos\theta) \equiv \sum_{L}\tilde{A}_L P_L(\cos\theta), \quad (13)$$

where the superscript $\mathrm{B}$ denotes quantities derived from long-range electron-dipole interactions calculated using the Born approximation. The term $\mathrm{d}\sigma^{\mathrm{B}}/\mathrm{d}\Omega$, representing the so-called Born closure [43] is incorporated into the coefficients $\tilde{A}_L$. For short-range potentials, the coefficients $A_L$ converge rapidly. In cases involving long-range electron-dipole interactions, where higher partial waves may be significant, employing the Born closure substantially reduces the required number of partial waves. Tests with the DCS of various polyatomic molecules at different energies showed that they can generally be fitted using the formula on the right-hand side in Eq. (13) with $L$=6, within a 95% confidence interval, provided that the DCS does not exhibit resonance-like structures.

The dashed line in Fig. 3 represents best fits of Eq. (13) with $L$=6 to the present experimental results. As shown, the fit results closely reproduce the measured values. Since the integral of the Legendre functions over the full angular range (0° to 180°) vanishes for $L \geq 1$, the TECS and its uncertainty are determined solely by the value of the coefficient $\tilde{A}_0$. The TECS derived from $\tilde{A}_0$ is presented in Table 2. The MTCS was calculated using all best-fit coefficients $\tilde{A}_L$, including those for $L \geq 1$.

Figure 4 (a) compares the experimental TECS with the results obtained using the POLYDCS code and the IAM-SCAR+I model, including rotational excitation cross sections. As expected, the ratio of the rotational excitation cross sections to the pure elastic scattering cross section decreases with increasing electron energy, starting at approximately 17% at 30 eV and decreasing to 6% at 800 eV. Overall, while the experimental TECS is slightly lower than the values predicted by the POLYDCS code, it aligns satisfactorily with the IAM-SCAR+I results (including rotational contributions) within experimental uncertainties. The TECS values reported by Hoshino et al. [9] are significantly lower than not only our results but also the predictions by the IAM-SCAR+I model and POLYDCS code, even at $T$ =100 eV, where both models are expected to deliver fairly accurate results. As shown in Fig. 3, this discrepancy primarily stems from the lower DCS values reported by Hoshino et al. [9], which propagate directly into their TECS calculations. Furthermore, additional differences may result from variations in the extrapolation techniques used for the angular integration from 0°-180°.

In Fig. 4 (a), the relative contributions of the molecular orbitals (MOs) associated with the irreducible representations (IR) $A_1$, $A_2$, and $E$ to $\sigma_{\mathrm{el}}$ are also displayed as a function of the electron energy $T$. To a first approximation, the relative contribution of each IR to the TECS corresponds to the ratio $r_{\mathrm{MO}}^{\mathrm{IR}}$, defined as the number of the MOs belonging to a specific IR divided by the total number of MOs. A noticeable deviation between these quantities is observed for the IRs $A_1$ and $A_2$. While the contribution of $A_1$ to the TECS is slightly lower than expected based on $r_{\mathrm{MO}}^{\mathrm{IR}}$ (33%), the contribution of $A_2$ is approximately twice as high (6%). This discrepancy is attributed to the characteristics of the MOs involved. Specifically, one of the two orbitals belonging to $A_2$ is the outermost HOMO, which significantly enhances its contribution to electron collision processes compared to $A_1$.

In addition to the measurement of elastic scattering cross sections, the TICS of $CCl_3F$ was calculated using the SCOP model (Fig. 4 (b)). The calculated values align well with the experimental total ionization cross sections reported by Sierra et al. [7] for $T \leq 80\ \mathrm{eV}$, remaining within the experimental uncertainties. It is important to note that, in the definition of the $\alpha$ parameter in the SCOP model, the threshold excitation energy was set equal to the ionization potential of $CCl_3F$. Thus, $\sigma_{\mathrm{inel}}$ corresponds to the total ionization cross section of $CCl_3F$.

Furthermore, the TICS was combined with the TECS and compared to the experimental TCS data from Jones [4] and Zecca et al. [5]. As shown in Fig. 4 (b), the sum of the TECS and TICS agree well with the experimental TCS $\sigma_{\mathrm{tot}}$ within uncertainties. Additionally, it can be seen from Fig. 4 that the TICS of $CCl_3F$ approach the TECS above 400 eV, reflecting the balanced contribution of elastic and inelastic scattering to the TCS at high energies - a phenomenon commonly observed in other molecules [16].

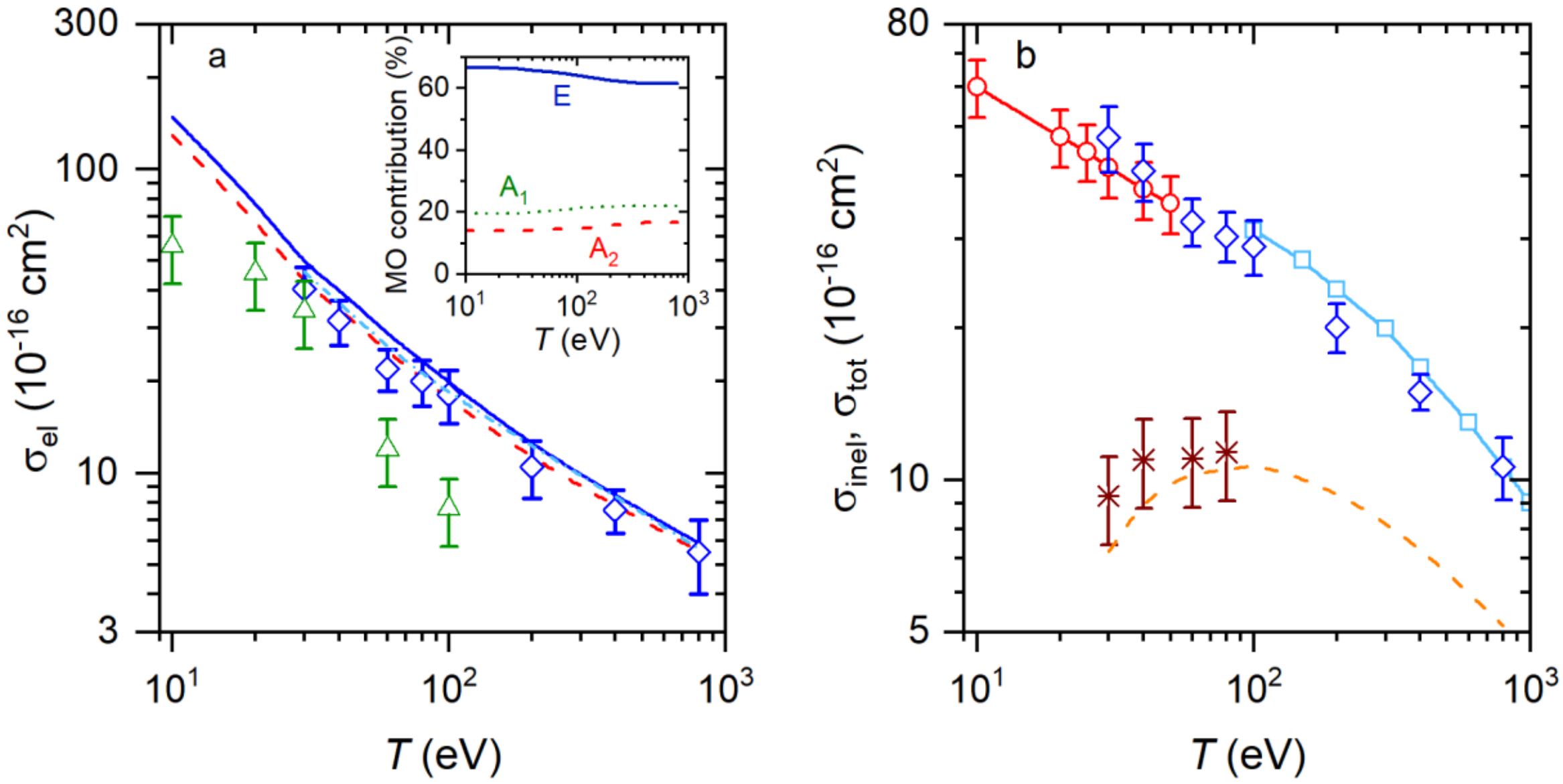

FIG. 4. **(a)** TECS ($\sigma_{\mathrm{el}}$) of $CCl_3F$ from the present work (◇) compared with theoretical and experimental results. Shown are values computed using the IAM-SCAR+I model including rotational contributions (– · –), and the close coupling approximation with the POLYDCS code excluding (– –) and including rotational contributions (—). Experimental data from Hoshino et al. (△) [9] are also included. The inset displays the contributions of the specific irreducible representations (IR) denoted by $\mathrm{E}$, $\mathrm{A_1}$, and $\mathrm{A_2}$ to $\sigma_{\mathrm{el}}$, as calculated using the POLYDCS code. **(b)** TICS ($\sigma_{\mathrm{inel}}$) of $CCl_3F$ calculated using the SCOP model (– –) are compared to the experimental total ionization cross sections reported by Sierra et al. (✳) [7]. The sum of $\sigma_{\mathrm{el}} + \sigma_{\mathrm{inel}} = \sigma_{\mathrm{tot}}$ determined in this work (◇) is compared to the TCS of $CCl_3F$ measured by Jones (○) [4] and Zecca et al. (□) [5].

## 6. Conclusions

The experimental results for the DCS of $CCl_3F$ agree satisfactorily with the IAM-SCAR+I model within experimental uncertainties for $T \geq 200\ \mathrm{eV}$. This is expected, as the IAM-SCAR+I model is known to reliably predict the DCS of polyatomic molecules at high electron energies. The poorest agreement was found in the angular range of 60° to 130° below 100 eV, where two pronounced minima were observed in the experimental data at electron energies. At 30 eV and 60 eV, differences of up to 100% were observed between the present measurements and the model predictions.

In this energy range, our results also show considerable deviations from the data of Hoshino et al., which are currently the only available experimental data. These deviations are observed not only in the absolute scale but also in the angular dependence. Notably, the data of Hoshino et al. closely match the angular dependence predicted by the IAM-SCAR+I model in contrast to our results. Total elastic scattering cross sections of $CCl_3F$, determined from the experimental DCS, are largely consistent with calculations performed with the IAM-SCAR+I model plus rotational excitation cross sections, within experimental uncertainties. However, they tend to be slightly lower than those obtained from the close coupling calculations with the POLYDCS code, which also includes rotational excitation cross sections. The latter contributes 17% to $\sigma_{\mathrm{el}}$ at 30 eV and decreases to 6% at 800 eV. Notably, the TECS reported by Hoshino et al. [9] are again lower than our experimental data and the predictions of the IAM-SCAR+I model.

Total inelastic scattering cross sections of $CCl_3F$, calculated using the SCOP model, reproduce the experimental total ionization cross sections available in literature satisfactorily within experimental uncertainties below 80 eV. Throughout the entire measured energy range, the sum of the TICS and TECS agrees well with the experimental TCS measured by other groups. This further indicates that the three types of the scattering cross sections $\sigma_{\mathrm{el}}$, $\sigma_{\mathrm{inel}}$, and $\sigma_{\mathrm{tot}}$ determined in the present work and by other groups are consistent with each other.

**Data Availability Statement** The experimental datasets generated and analyzed during the current study, including the $N_2$ reference data, are available in Zenodo [44].

**Acknowledgement** This research was supported by the joint research project BIOSPHERE. The project 21GRD02 BIOSPHERE has received funding from the European Partnership on Metrology, co-financed by the European Union's Horizon Europe Research and Innovation Programme and by the Participating States. The authors express their thanks to Heike Nittmann and Andreas Pausewang for their assistance and technical support during the measurements. The authors also thank Felix Lehner for his technical support while setting up the VOLSCAT and POLYDCS code and calculations.